\documentclass[aps,preprint,prd,showpacs,nofootinbib]{revtex4}
\usepackage{amsmath}
\usepackage{graphicx}
\usepackage{dcolumn}
\usepackage{bm}
\usepackage{amssymb}
\usepackage{latexsym}

\def\be{\begin{equation}}
\def\ee{\end{equation}}
\def\ba{\begin{eqnarray}}
\def\ea{\end{eqnarray}}

\begin{document}

\title{Can Universe Experience Many Cycles with Different Vacua ? }
\author{Yun-Song Piao$^{a,b}$}
%\email{yspiao@itp.ac.cn}
\affiliation{${}^a$Institute of High
Energy Physics, Chinese Academy of Science, P.O. Box 918-4,
Beijing 100039, P. R. China} \affiliation{${}^b$Interdisciplinary
Center of Theoretical Studies, Chinese Academy of Sciences, P.O.
Box 2735, Beijing 100080, China}

\begin{abstract}

%The facts from observation and theory have lead an increased
%attention to the notion that the number of vacuum states may be
%enormous,
Recently, the notion that the number of vacua is enormous has
received increased attentions, which may be regarded as a possible
anthropical explanation to incredible small cosmological constant.
Further, a dynamical mechanisms to implement this possibility is
required. We show in an operable model of cyclic universe that the
universe can experience many cycles with different vacua, which is
a generic behavior independent of the details of the model. This
might provide a distinct dynamical approach to an anthropically
favorable vacuum.

\end{abstract}

\pacs{98.80.Cq, 98.70.Vc} \maketitle

Recently, the notion that the number of vacua is astronomical has
received increased attentions, which may attribute to two
developments of observations and theory. The first is the
accelerated expansion of present universe implied by supernovae
SNIa \cite{PR}, to which the simplest explanation is a small
cosmological constant. However, its incredible smallness and fine
tuning make us hard to find a vacuum in a range of observed value
unless there are an enormous of solutions with almost all possible
values. The second is that it is known that there are a large
number of vacuum states in string theory \cite{Douglas, KKLT}, see
also Ref. \cite{BP, FMSW}. Recently the space of all such vacua
has been dubbed landscape \cite{Susskind}. In some sense, the low
energy properties of string theory can be approximated by field
theory. Thus the landscape can also be described as the space of a
set of fields with a complicated and rugged potential, where the
local minima of potential are called the vacua. When this local
minimum is an absolute minimum, the vacuum is stable, and
otherwise it is meta stable. The value of potential energy at the
minimum may be regarded as the cosmological constant for that
vacuum. Since the number of vacua is exponentially large,
%the tunnelling
%between various vacua can be regarded as an explanation to a small
%cosmological constant observed \cite{BP, FMSW, KKLT},
there can be enough vacua with small cosmological constant. The
probability that an observer find a cosmological constant no
greater than the value observed will be not too small, which is
often referred as an interesting application for anthropic
principle \cite{W, MSW}, see Ref. \cite{Smolin} for a recent
comment.

Though string theory brings us diverse vacua, which makes us
possible to solve the problem of cosmological constant puzzling us
for a long time, we still require a dynamical mechanism to
implement this possibility. Eternal inflation \cite{VL} based on
inflation scenario \cite{G, LAS} is often regarded as a natural
candidate, where an observer starts with a large value of vacuum,
with bubble nucleating over potential barrier he will see a series
of vacuum descending. The chances that he lands in an observable
vacuum may be small, but still acceptable when we consider
possible infinite numbers of bubbles. But the realistic evolution
of universe controlled by a landscape might show itself more
complex than that expected, which may spur our more thinks. Note
that recently, the cyclic model has been proposed as a radical
alternative to inflation scenario \cite{KOS}, which is motivated
by the string/M theory, where the relevant dynamics can be
described by an effective field theory. The separation of the
branes in the extra dimensions is modeled as a scalar field. The
idea that universe is cyclic is ancient, which can be found in
mythologies and philosophies dating back to the beginning of
recorded history of mankind, and has been still an aesthetic
attraction. Thus with diverse environments which string theory is
likely to bring us, a naive question is whether the universe can
experience many cycles with different vacua.

In string landscape there are a large number of dS minima as well
as AdS minima. The cosmology of scalar field with AdS minimum has
been investigated in Ref. \cite{FFKL} extensively. It is well
known that for an expanding universe with dS minimum, it will
arrive at a dS regime asymptotically. However, the universe with
AdS minimum can not approach an AdS regime, and instead it will
stop expanding and enter into a collapsing phase. During the
collapsing process, its energy goes up rapidly. The singularity
will become inevitable unless there is a mechanism responsible for
the bounce. It has been noticed \cite{KSS} that when the universe
is in the contracting phase, the field can be driven and roll up
along its potential, and arrive at a large enough value for a
successful inflation to occur in expanding phase \cite{PFZ}.
%which can be implemented by a feasible bounce .
These pioneer attempts have given us a partial answer. In the
following we will focus on a simple example that allows us to
obtain our basic conclusion.
% and give further explanations.
We show that the universe can experience many cycles with
different vacua and further explain how a cycle with an early
successful inflation and a small cosmological constant may be
anthropically selected as the universe in which we live.
%leave a systematic search in a coming
%publication.
%In this letter, we study the cosmological phenomena of
%scalar field potential with multi AdS minima. We find that for a
%supposed bounce the field can be driven and stride over some hills
%and valleys of potential from a AdS minimum to another during
%each contraction/expansion cycle of universe, which is a generic
%behavior and may has an interesting imply in the landscape of
%string theory. We take a landscape of two dimension as example and
%discuss a distinct implement of anthropic principle,

Let us start with such an effective Lagrangian of single scalar
field as follows \be {\cal L} = {1\over 2}
\partial_{\mu} \varphi
\partial^{\mu} \varphi -\Lambda_*\left(1-\cos({m\over \sqrt{\Lambda_*}}\varphi)\right)+\Lambda\ee
where $\Lambda$ is a small positive constant which makes the
minima of periodic potential AdS's. The potential is plotted in
the left-up of Fig. 1.

%When $| {m\over \sqrt{\Lambda_*}}\varphi -2n\pi | \ll 1$, where
%$n$ is an integer, the potential is ${1\over 2}(m \varphi -2n\pi
%\sqrt{\Lambda_*})^2 -\Lambda$.

No losing generality, we take the field around the origine as
example, where the potential can be written as $ {1\over 2} m^2
\varphi^2 -\Lambda$. We assume that initially the universe is in a
contracting phase and the field $\varphi$ is in $\varphi \simeq 0$
minima of its potential and has a small ${\dot \varphi}$
responsible for $\rho_{\varphi} >0$.
%oscillate near $\varphi=0$.
In this case $m^2\gg h^2 \equiv ({{\dot a}\over a})^2\sim
\rho_{\varphi}$, where $8\pi /m_p^2 =1$ has been set,
%{\it i.e.}
%the frequency of oscillation is much larger than the evolution
%rate of universe,
the term $3h{\dot \varphi}$ can be neglected. Thus the equation of
motion of $\varphi$ can be reduced to \be {\ddot
\varphi}+m^2\varphi \simeq 0 \label{dphi1}\ee which can be solved
as an oscillation with frequency $m$, $ \varphi \simeq \varphi_{a}
\sin{(mt)} \label{ophi}$, where $\varphi_{a}$ is the amplitude of
the oscillation.
%Instituting Eq. (\ref{ophi}) into
%Eq. (\ref{p}), we have \be p_{\varphi} \simeq {m^2 \varphi_{\rm
%am}^2\over 2} \cos{(2mt)}\ee
When taking the time average over many oscillations of field,
$<p_{\varphi}>\simeq 0$ is obtained. Thus around the minimum of
the field the behavior of the universe is similar to the case
dominated by usual matter.
%Thus in this regime the universe
%contracts as $a \sim t^{2/3}$ and the energy density of the field
%grows as $\rho_{\varphi} \sim a^{-3}$. From Eqs. (\ref{p}) and
%(\ref{ophi}), we have $\rho_{\varphi}\sim m^2\varphi_{\rm am}^2$,
%thus the amplitude of oscillation increases as $\varphi_{\rm
%am}\sim a^{-3/2}$.

\begin{figure}[t]
\begin{center}
\includegraphics[width=8cm]{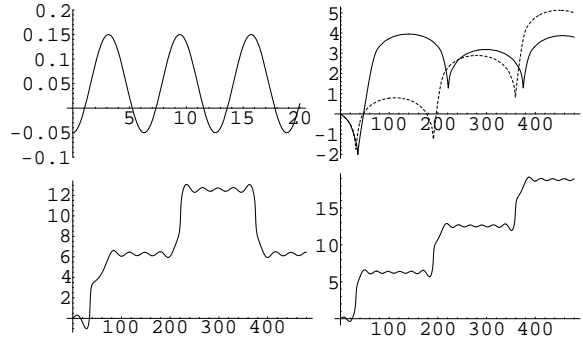}
\caption{ The upper left panel is the figure of potential with
$\Lambda_* =0.1$, ${m\over \sqrt{\Lambda_*}}=1$ and
$\Lambda=0.05$. The upper right panel is the figure of $\ln{a}$
with respect to time. The dashed and solid line are the cases that
the hight $\Lambda_*$ of potential are taken as $0.07, 0.04$,
respectively, where $\Lambda=0.005$ is taken. The lower panel are
the figures of field with respect to time, where $\sigma =1$ is
taken. We find that for different cycles of universe the field
oscillates in different minima. }
\end{center}
\end{figure}

In a collapsing universe, $3h{\dot \varphi}$ is anti-frictional.
Instead of damping the motion of $\varphi$ in the expansion, it
accelerates the motion of $\varphi$ and makes the energy of field
increase. Since $h\simeq m \varphi_{a} $ and ${\dot \varphi}\simeq
m\varphi$, when $\varphi\gtrsim 1$, we have
$3h{\dot\varphi}\gtrsim m^2\varphi$. Thus with the increase of
amplitude of oscillating field, when the term $3h{\dot \varphi} $
can not be neglected any more, the oscillation of field ends and
the universe will enter into the regime dominated by kinetic
energy of field rapidly, $ {\dot \varphi}^2 \gg m^2\varphi^2 $,
{\it i.e.} the detail of potential has been not important, thus
$p_{\varphi}\simeq \rho_{\varphi}$. Instead of Eq. (\ref{dphi1}),
\be {\ddot \varphi}+3h{\dot \varphi}\simeq 0 . \label{kphi}\ee can
be satisfied.
%From Eq. (\ref{kphi}), we have \be
%{\dot \varphi} \simeq {c\over a^3} \label{phia}\ee where $c$ is
%the integral constant.
This leads to $\rho_{\varphi} \sim {\dot \varphi}^2 \sim 1/a^6 $,
thus we obtain \be a^3\simeq c\sqrt{3\over 2}(t_{s}-t)
\label{a3s}\ee where $c$ is an integral constant and $t_{s}$ is
the time when $a\simeq 0$, which means that the universe will
collapse into singularity unless the bounce occurs, and \be
\varphi \simeq \varphi_{k} + \sqrt{2\over 3}\ln{({t_{s}-t_{
k}\over t_{s}- t})} \label{phibou}\ee where $\varphi_{k}$ and
$t_{k}$ are the values of field and time just entering into the
regime dominated by kinetic energy, respectively.

We expect that Big Crunch singularity is not a possible feature of
quantum gravity and there might be some mechanism from high
energy/dimension theories responsible for a non-singular bounce.
We suppose, following this line, that in high energy regime the
Friedmann equation can be modified as
%where the observable universe is embedded in high dimension
%space/time, see Ref. \cite{M} for a recent introduction. The
%cosmological constant of brane has been offset by that of bulk.
%The $\sigma$ is the brane tension and the numerical factor has
%been absorbed into $\sigma$. In the meantime the dark radiation
%and the curvature terms are neglected,
$ 3 h^2 \simeq \rho_{\varphi} -\rho_{\varphi}^2 / \sigma$
\cite{SS},
%$\rho_{\varphi}$ is the energy density of
%the field.
%Further, it
%has been pointed out in Ref. \cite{CF} that due to the effects of
%the bulk and Israel matching conditions any relationship between
%$h$ and $\rho_{\varphi}$ is possible.
%This is an only assumption made in this report.
%Regardless the Eq.
%(\ref{h1}) can be or not be regarded as a possible input from a
%realistic theory of quantum gravity,
which is only required to carry out visual numerical calculations.
The final results are independent on this detail that the bounce
is implemented.
% and in general is not standard
%Fridmann-like.
During the contraction, the field will be driven up along its
potential. When it arrives at apex of potential hill, it will roll
down to another minimum. Due to the endless growing of its kinetic
energy, in some time the modification term of Friedmann equation
will become important, and when $\rho \simeq \sigma$, we obtain
$h\simeq 0$, thus the collapse of universe will be hold back and
then enter into an expanding phase. The behavior of scale factor
around the bounce can be solved approximately as $ a^6 \simeq 3
c^2 (t-t_{b})^2/2 + c^2/ 2\sigma$ where $t_{b}$ is
%the time when the bounce occurs and is the
an integral constant.
%determined by $\sigma$ as well as the value
%of ${\dot \varphi}$ and $a$ at $t\simeq t_{\rm kin}$.
When $t\simeq t_{b}$, the scale factor of universe arrives its
minimum.

\begin{figure}[]
\begin{center}
\includegraphics[width=8cm]{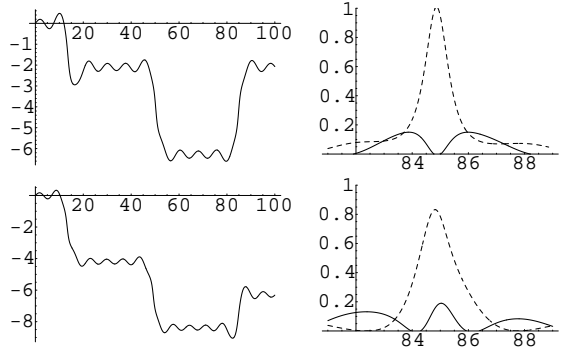}
\caption{ The left side panels are the figures of field with
respect to time, while the right side panels are the figures of
its potential energy (solid line) and kinetic energy (dashing
line) in time interval (81, 89). The upper right panel is the case
that the field strides over two hills and one valley and the lower
right panel is the case of one hill and one valley.   }
\end{center}
\end{figure}

The universe, controlled by multi AdS minima and a bounce
mechanism in high energy, will show itself many oscillations. We
plot the functions of the field and scale factor with respect to
time in Fig. 1, where when the modification scale $\sigma$ is
large than the height of potential hill, the field will be driven
from a minimum to another, and in each cycle of oscillating
universe, the field will generally lie in different minima
\cite{PZhang}.
%Which minimum is selected is
%dependent on the shape of potential, such as $\Lambda_*$ and $m$.

%\begin{figure}[t]
%\begin{center}
%\includegraphics[width=8cm]{da.eps}
%\caption{The scale factor $\ln{a}$ as a function of time. The long
%and short dashing, and solid line are the cases that the hight
%$\Lambda_*$ of potential are taken as $0.1, 0.07, 0.04$,
%respectively. }
%\end{center}
%\end{figure}

%When $m^2 \varphi_{k}^2 \gtrsim \Lambda$,
The value to which the field is driven during each
contraction/expansion cycle can be simply estimated. When the
field just enters into the regime dominated by kinetic energy,
${\dot \varphi}_{k}^2\simeq m^2\varphi^2_{k}$ can be satisfied
approximately, and when the field is about at the point of the
bounce, its value is ${\dot \varphi}_{b}^2 \simeq \sigma$. Thus
from Eq. (\ref{a3s}), \be \left({t_{s}- t_{b}\over
t_{s}-t_{k}}\right)^2 = \left({a_{b}\over a_{k}}\right)^6 =
\left({{\dot \varphi}_{k}\over {\dot \varphi}_{b}}\right)^2\simeq
{m^2 \varphi_{k}^2\over \sigma}\ee can be obtained. When it is
substituted into Eq. (\ref{phibou}), the maximum value is given by
\be \varphi \simeq\varphi_{k} + \sqrt{2\over 3}\ln{({\sigma\over
m^2 \varphi_{ k}^2})}\label{phibou2}\ee where the second term has
been doubled since during kinetic dominated period the behavior of
the field can be regarded approximately as symmetrical before and
after the bounce, which can also be seen from figures of right
side of Fig. 2. In general $\varphi_{k}\simeq 1$, thus according
to Eq. (\ref{phibou2}), the change of field can be written as \be
1 + {1\over 2}\sqrt{2\over 3}\ln{({\sigma\over m^2 })}\lesssim
\triangle\varphi \lesssim 1 + \sqrt{2\over 3}\ln{({\sigma\over m^2
})} \label{detphi}\ee
%For a reasonable estimation, we take $\varphi_{\rm ki}\sim 1$,
%thus obtain $\varphi_{\rm rev}\lesssim 4$ for $\sigma =0.0001$ and
%$\varphi_{\rm rev}\lesssim 10$ for $\sigma =0.1$, which is
%compatible with the numerical results of Fig. 4. In this case
%there is a up-limit $\varphi_{\rm rev}\lesssim 11$ for $\sigma
%=1$, in which the modification of Friedmann Eq. is at Planck
%scale.
%The maximal value of the changing of field during each
%contraction/expansion cycle can
which is only determined by the mass $m$ around its minima and the
modification scale $\sigma$. In Fig. 1, $\sigma =1$ and $m^2=0.07,
0.04$. From Eq. (\ref{detphi}), $2.5 <\triangle\varphi <3.5$ is
obtained, which makes the field just stride over a hill during
each cycle and can be seen from Fig. 1, where the distance between
the apex of hill and the valley is 3.14.

%For the universe
%initially in a state with AdS minimum \footnote{This may be from
%the rolling from positive potential or the tunnelling from some dS
%minimum. The universe tunnelling into a state with AdS minimum
%will generically undergo a Crunch \cite{Banks}.
%It has been pointed out in Ref. \cite{BDG} that while
%the jumping between dS states might make sense in the
%semiclassical approximation, the jumping to regions with AdS
%minimum always involves a breakdown of effective field theory
%\cite{Banks}.
%}, it will eventually crunches, and in the meantime its energy
%increases rapidly. In some time the mechanism expected and
%breaking the effective theory will hold back this crunch and make
%the universe reverse. During this process the field describing the
%space of vacua can be driven to its neighboring or more far hills
%or valleys. Dependent on the detail of landscape, the field will
%roll down and arrive at eventually some AdS or dS minimum
%\footnote{ }.
%If one think that in a consistent model of quantum gravity the
%universe can not evolve into a Big Crunch, such a system with many
%AdS and dS minima, where any meta-stable dS minimum can tunnel to
%any other and further one of them can tunnel to a Crunch, might be
%inconsistent.

Further, these results can be straightly applied to the case with
multi degrees of scalar freedom, whose potential in field space is
reflected in a complicated terrain in real space. In Fig. 3, the
potential is plotted as a function of two scalar fields, where
there are many hills and valleys with AdS and dS minima
\cite{Giddings}. The similar behaviors can be found. Initially the
fields are in some valley with AdS minimum of landscape and
oscillate \cite{Banks}. When the energy of oscillation can be
compared with this minimum, the universe will stop expanding and
enter into a collapsing phase. During the contraction, the kinetic
energy of the fields will rise rapidly and be far larger than
their potential energy, which can be seen from the dashed line of
Fig. 3. When the total energy arrives the modification scale,
$h\simeq 0$, the universe will return and expand. With the
expansion of universe, the kinetic energy of fields decreases
rapidly. Finally the fields will be ``dropped" into some hill or
valley, which is generically different from previous one. This
scenario can be visualised as that an airplane takes off from a
valley, and as close to cloud of modification scale, it returns
and lands on some hill \cite{BDG} or in some other valley, see
Fig. 3. During this period the number of hills and valleys it
flies over is determined by the detail of the landscape and the
bounce scale. For the case that the landscape has more AdS minima
than dS's, or even if the field can stay in a positive potential
temporarily it will also eventually roll to minimum with negative
potential (for further illustration see Fig. 4), periodical
fly/landing process will be inevitable and highly effective. In
each cycle, the universe lies generically in different states.
%As long as the universe enters into a state with AdS minimum,
%which may come from the tunnelling from dS minima or the landing
%of last fly, such fly will be inevitable.

\begin{figure}[t]
\begin{center}
\includegraphics[width=8cm]{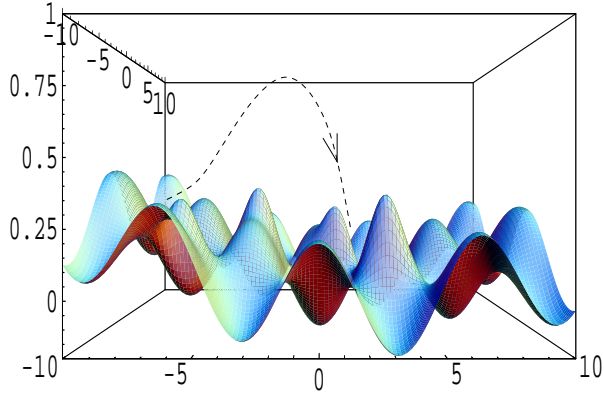}
\caption{ The 2D landscape. The potential is plotted as a function
of two scalar fields. The dashed line is the evolution of total
kinetic energy of fields during some cycle. }
\end{center}
\end{figure}

\begin{figure}[t]
\begin{center}
\includegraphics[width=8cm]{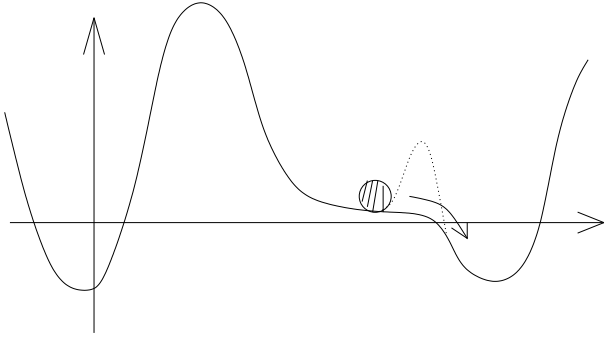}
\caption{ When the minima of valleys which the field rolls to
finally are dS minima, the cycle will generally cease, (or occur
in subsequently nucleated bubbles). To make universe cycle more
effectively, the potential (solid line) in the figure is a
feasible example. In almost all cases the field enters into
negative potential not by tunnelling over hill (dashed line) but
by rolling, which is locally just the case in cyclic model
\cite{KST}. But different from it, here the universe will be in
different regions of potential of fields in each cycle. Not in all
cycles are there such local potentials making field show a
quintessence-like behavior at present as in cyclic model
\cite{KOS}. However, when the number of cycles is large or
approaches infinity, some cycles similar to the universe which we
live in will be possible. }
\end{center}
\end{figure}

The quantum and thermal fluctuations will increase during the
contraction of each cycle. However, the increase of their energy
is less fast than that of the field energy $\sim 1/a^6$. Thus as
long as initially the fluctuations energy is lower than the
background energy, it will not exceed forever. The form of black
holes may be a problem \cite{Blackhole}, and further black hole
gas accumulated \cite{BF} after some cycles may make the universe
cease to cycle. However, there is a possibility that the field
could land on some potential hill or nearly flat plain after one
or several cycles. In this case there is a large room for
inflation to occur, which may dilute these leftovers not required
and seed primordial perturbations responsible for structures of
observable universe. Since the landscape is a complex function
with many scalar fields (degrees of freedom), from a hill to a
bottom of a valley, a large number of paths can be expected to
implement inflation.
%\footnote{ From Eq.
%(\ref{detphi}), for some suitable values of $m$ and $\sigma$,
%$\triangle\varphi \sim 1-10~ m_p$, see also Ref. \cite{PZhang},
%which is suitable for a successful inflationary cosmology but is
%far away from eternal inflation \cite{L2}.}.
It has been pointed out \cite{Easther} that in this case inflation
can be driven by several distinct fields, but the rolling of only
one field is dominant in each segment of path. When the minimum of
valley which the field rolls to finally is AdS's,
%(or
%even though the minimum is dS's, after one or many times
%tunnelling, the field will be also in AdS regime,)
the universe will collapse eventually and begin to the next cycle.
Such a cycle (fly, landing then (not) inflation) can occur time
after time.
%In each fly, the universe experiences a
%contraction/expansion cycle and selects different vacua depicting
%different low energy physics.
When the number of cycles is large or approaches infinity,
whichever minimum initially the universe is in, it can run over
almost all vacua of the landscape. Thus for a given landscape if
there are some vacua suitable to us, there are certainly some
cycles where we can live.
%In other words, if the universe is dominated by
%a certain united theory, one of whose solutions is interesting to
%us, there are certainly some cycles with this solution in all
%cycles that the universe experiences.
In some extent our universe with an early successful inflation and
a late time small cosmological constant may be regarded as an
anthropic one of many cycles.

In conclusion, we show that the universe can experience many
cycles with different vacua, which may be regarded as a feasible
approach to an anthropically favorable vacuum. In our example the
universe is controlled by a periodic scalar field potential with
AdS minima. We find that during each contraction/expansion cycle
of universe, the field will be driven from a minimum to another.
The usual anthropic proposal for small observed cosmological
constant assumes that there are many universes, or in the universe
there are many different regions based on eternal inflation
\cite{LLM, V}, while in our proposal, the universe experiences
many cycles and in each cycle it lies in different vacua. It is in
this sense that it might be another interesting dynamical
implement different from eternal inflation to solve the problem of
cosmological constant.
%However, the realistic landscape may be more
%complex, a possible case is that both mechanisms appear
%simultaneously.
%However, in this report, we shall discuss a distinct
%anthropic picture, where there are many cycles(periods) for an
%(maybe multi) universe, each cycle has different vacua, and the
%existence of life selects certain ones.
We in this paper only simply used a known modification of
Friedmann equation to avoid Big Crunch. Though our study might be
idealistic, we believe that we have identified the basic
ingredient of the required answer. A more realistic bounce
mechanism expected from possible quantum gravity theory is
required, which is also significant to avoid a potentially
catastrophic instability of the landscape with meta-stable dS
minima \cite{BDG}. Some further issues, especially how to embed it
into string/M theory, and interesting applications is left in
future works.

%Though this pastime is excessively simple and idealised and looks
%an awkward game, it might have a little helps for understanding
%some things.
%The increasing of perturbations and
%anisotropies and the inclusion of other light scalar fields and
%matter are not considered in detail.
%Further, how to embed this toy scenario in a realistic string/M
%theory is an important issue.
%To implement them may not be easy
%and has gone beyond the scope of this brief report.
%We might return these aspects in the future.

{\bf Acknowledgments} The author would like to thank Miao Li, Yi
Ling for helpful discussions and Mingzhe Li for valuable comments.
This work is supported by K.C. Wang Postdoc Foundation.

%This work is supported by K.C.Wang
%Postdoc Foundation.


\begin{thebibliography}{99}

\bibitem{PR} S. Perlmutter, et al. Ap. J. \textbf{483}, 565 (1997);
A.G. Riess, et al. Astron. J. \textbf{116}, 1009 (1998).


\bibitem{Douglas} M.R. Douglas, JHEP \textbf{0305} 046 (2003).

\bibitem{KKLT} S. Kachru, R. Kallosh, A. Linde and S.P. Trivedi, Phys. Rev. \textbf{D68}
046005 (2003).

\bibitem{BP} R. Bousso and J. Polchinski, JHEP \textbf{0006}, 006
(2000).

\bibitem{FMSW} J.L. Feng, J. March-Russell, S. Sethi and F.
Wilczek, Nucl. Phys. \textbf{B602}, 307 (2001).

\bibitem{Susskind} L. Susskind, hep-th/0302219; B. Freivogel and
L. Susskind, hep-th/0408133.

\bibitem{W} S. Weinberg, Phys. Rev. Lett. 59, 2607 (1987);
Phys. Rev. \textbf{D61} 103505 (2000); astro-ph/0005206.

\bibitem{MSW} H. Martel, P. Shapiro and S. Weinberg, Astrophys. J.
\textbf{492}, 29 (1998).

\bibitem{Smolin} L. Smolin, hep-th/0407213.

\bibitem{VL} A. Vilenkin, Phys. Rev. \textbf{D27} 2848 (1983); A.
Linde, Phys. Lett. \textbf{B175} 395 1986, see also A. Guth, Phys.
Rept. \textbf{333} 555 (2000).

\bibitem{G} A.H. Guth, Phys. Rev. \textbf{D23} (1981) 347;

\bibitem{LAS} A.D. Linde, Phys. Lett. \textbf{B108} (1982) 389; A.A. Albrecht
and P.J. Steinhardt, Phys. Rev. Lett. \textbf{48} (1982) 1220.



\bibitem{KOS} J. Khoury, B.A. Ovrut, P.J. Steinhardt and N. Turok,
Phys. Rev. \textbf{D64} (2001) 123522; P.J. Steinhardt and N.
Turok, Science \textbf{296}, (2002) 1436; Phys. Rev. \textbf{D65}
126003 (2002); see also P.J. Steinhardt and N. Turok,
astro-ph/0404480 for a recent review.

\bibitem{FFKL} G. Felder, A. Frolov, L. Kofman and A. Linde, Phys.
Rev. \textbf{D66} 023507 (2002).

\bibitem{KSS} N. Kanekar, V. Sahni and Shtanov, Phys. Rev.
\textbf{D63} (2001) 084520.



\bibitem{PFZ} This has been used to provide a
possible explanation for low CMB anisotropies on large angular
scale, see Y.S. Piao, B. Feng and X. Zhang, Phys. Rev.
\textbf{D69} 103520 (2004).


\bibitem{SS} This may be
motivated in a brane world scenario, see Y. Shtanov and V. Sahni,
Phys. Lett. \textbf{B557} 1 (2003).

\bibitem{PZhang} Further, some details of bouncing and oscillating universe with a
massive scalar field and a AdS minimum can be found in Ref. Y.S.
Piao and Y.Z. Zhang, gr-qc/0407027.







\bibitem{Giddings} For the case that the hills and valleys of
landscape roll off into a flat plain, the decompactification of
extra dimension will occur and the universe evolves into higher
dimensions, see S.B. Giddings, Phys. Rev. \textbf{D68}, 026006
(2003); S.B. Giddings and R.C. Myers, hep-th/0404220.




\bibitem{Banks} This may be
from the rolling from positive potential or the tunnelling from
some dS minima. The latter maight or might not occur. If possible,
the universe tunnelling into a state with AdS minimum will
generically undergo a Crunch, see T. Banks, hep-th/0211160;
hep-th/0306074.


\bibitem{BDG} In some sense, this can be regarded as a excitation
from AdS minima to meta-stable dS states. In fact it has been
noticed that this possibility depends precisely on the fact that
the universe evolves from a Big Bang singularity. This is because
the increase of energy density can make the universe go over the
barrier to other minima, see T. Banks, M. Dine and E. Gorbatov,
hep-th/0309170.




\bibitem{KST}  J. Khoury, P.J. Steinhardt and N. Turok, Phys. Rev. Lett. \textbf{92}
031302 (2004).

\bibitem{Blackhole} This may add the entropy of each cycle, which
is not considered in our numerical calculations for simplicity.

\bibitem{BF} T. Banks and W. Fischler, hep-th/0212113.

%\bibitem{L2} A.D. Linde, Phys. Lett. \textbf{B175} 395 (1986).

\bibitem{Easther} R. Easther, hep-th/0407042.

\bibitem{LLM} A. Linde, D. Linde and A. Mezhlumian, Phys. Rev.
\textbf{D49} 1783 (1994).

\bibitem{V} A. Vilenkin, Phys. Rev. Lett. \textbf{74} 846 (1995);
J. Garriga and A. Vilenkin, Phys. Rev. \textbf{D64} 023507 (2001).


\end{thebibliography}
\end{document}